\documentclass[aps,pre,amsmath,amssymb,twocolumn]{revtex4}
\usepackage{amsmath,amsthm,amssymb}
\allowdisplaybreaks
\usepackage{graphicx}
\usepackage{epstopdf}
\usepackage{pstricks}
\usepackage{epsf}
\usepackage{xcolor}

\numberwithin{equation}{section}

\newcommand{\be}{\begin{equation}}
\newcommand{\ee}{\end{equation}}
\newcommand{\ba}{\begin{eqnarray}}
\newcommand{\ea}{\end{eqnarray}}
\newcommand{\baa}{\begin{eqnarray}}
\newcommand{\eaa}{\end{eqnarray}}
\newcommand{\ed}{\end{document}}
\newcommand{\lab}[1]{\label{#1}}
\newcommand{\re}[1]{(\ref{#1})}
\newcommand{\ci}[1]{\cite{#1}}

\begin{document}

\title {Exciton dynamics in branched conducting polymers: Quantum graphs based approach}
\author{J.R. Yusupov$^a$, Kh.Sh. Matyokubov$^b$, K.K. Sabirov$^c$  and D.U. Matrasulov$^a$}
\affiliation{ $^a$ Turin Polytechnic University in Tashkent, 17
Niyazov Str., 100095,  Tashkent, Uzbekistan\\
$^b$ Urgench State University, 14 H. Olimjon Str., 220100,
Urgench, Uzbekistan\\
$^c$Tashkent University of Information Technologies, 108 Amir
Temur Str., 100200, Tashkent Uzbekistan}

\begin{abstract}
We consider dynamics of excitons in branched conducting polymers. An effective
model based on the use of quantum graph concept is applied for computing of
exciton migration along the branched polymer chain. Condition for the regime,
when the transmission of exciton through the branching point is reflectionless
is revealed.
\end{abstract}
\maketitle

\section{Introduction}

Modeling the charge carrier migration in conducting polymers is of practical
importance for material and device optimization in organic electronics.
Effective functionalization of conducting polymers for different purposes
requires understanding of the mechanisms for charge transport and their
utilization. So far, considerable progress has been made in the developing of
different models for charge carrying quasiparticles in conducting polymers
\ci{MGH,Wallace,Kumar,Salaneck,Heeger1,Heeger2,Heeger3,Chsol1,Chsol2,Chsol3,Chsol4,Chsol5,Chsol6,SSH,TDFT,PPP,TDHF},
such as polarons, excitons and solitons. Excitons, which are the bound
electron-hole pair states, appear, e.g., in photovoltaic processes in
conducting polymers interacting with optical field. Charged solitons in
conducting polymers provide another mechanism for charge transport. When charge
is trapped in the phonon cloud by forming a bound state, they are called
polarons. Each mechanism for charge carrier transport plays important role
depending on the type of the functionalization. Therefore, effective
utilization of these mechanisms in organic electronics requires developing
different effective models for charge carrier dynamics. Among these three
mechanisms of charge transport, exciton mechanism is of importance for
practical applications in organic photovoltaics and organic electronics. Tight
binding theory of excitons in conducting polymers was proposed in \ci{Rice}.
Different aspects of excitons, including band structure calculations, field
induced ionization, continuum approach for exciton transport and optical
properties of conducting polymers have been studied in the
Refs.\ci{Braz1,Braz02,Braz2,Braz3,Braz4}. In \ci{Kobrak} dynamic model of
self-trapped excitons is studied using the Schr\"odinger equation on polymer
lattice by taking into account exciton-phonon interaction. Different aspects of
exciton dynamics, such as ultrafast exciton dissociation \ci{Bittner1}, exciton
dissociation at donor-acceptor polymer heterojunctions \ci{Bittner2}, exciton
diffusion \ci{Bittner3} have been studied by E.~Bittner and co-workers.

Most of the conducting polymers synthesized so far have linear (unbranched)
structure.  However, conducting polymers having branched architecture attracted
much attention recently (see, e.g.
Refs.\ci{BCP1,BCP2,BCP3,BCP5,BCP6,BCP7,BCP8,BCP9,BCP10,BCP11,BCP12,BCP13,BCP14,BCP15,BCP16}).
These are kinds of polymers, in which a linear chain splits into the two or
more branches starting from some point, which is called branching point, or
node. The structure of a branching can have different architecture, e.g. can be
in the form of star, tree, ring, etc. These latter implies the rule for
branching and called branching topology. When the topology of a polymer is very
complicated, it is called hyperbranched polymer. Branched polymers differ from
their linear counterparts in several important aspects. Such polymer forms a
more compact coil than a linear polymer with the same molecular weight. Also,
depending on the topology of branching, electronic and elasticity properties
can be completely different than those of linear polymers. Despite the
considerable progress made in the synthesis and study of branched conducting
polymers, the problem of the charge carrier dynamics in such structures is
still remaining as less studied topic. The only model for the study of exciton
dynamics in branched conducting polymers and other quasi-one-dimensional
molecular chains, to our knowledge, was proposed in a series of papers by
V.~Chernyak and co-authors in the
Refs.\ci{Chernyak31,Chernyak34,Chernyak35,Chernyak36,Chernyak37,Chernyak39,Chernyak42,Chernyak43},
where so-called exciton scattering method was developed. The approach considers
branched conjugated polymers as graphs and allows to construct scattering
matrices, describing exciton scattering at the vertices. One of the advantages
of the model proposed first in the Ref.\ci{Chernyak31} is the fact that it
allows to describe multi-exciton case. Later, in the
Refs.\ci{Chernyak34,Chernyak36} the approach was applied to compute exciton
scattering characteristics and excitation energies in different conjugated
polymers and quasi-one-dimensional molecules. The main assumption of the
exciton scattering model was small sizes of the exciton compared to the length
of the polymer branch and considering exciton as the standing wave. In such
approach, the method gives the results, which are in good agreement with
quantum chemical computations.

In this paper we consider dynamics of excitons in branched conducting polymers
by modeling them in terms of so-called quantum graphs. These latter are the
system of quantum wires connected to each other according to some rule, which
is called topology of a graph.  Linear, i.e. unbranched counterpart of such
polymers have been extensively studied earlier in the literature within the
different approaches (see, e.g., Refs.\ci{Heeger1,Heeger2,Heeger3}). The model
we use is based on the linear Schr\"odinger equation on metric graphs. Apart
from branched polymers, quantum graphs have been extensively studied earlier in
different contexts
\ci{Kost,Uzy1,Kuchment04,Uzy2,Exner15,Keating,Uzy3,KarimBdG,PTSQGR,Jambul,Jambul02}.
For modeling of exciton migration and possible reflectionless transmission of
excitons through the polymer branching point, we use so-called concept of
transparent boundary conditions for quantum graphs proposed recently in the
Refs.\ci{Jambul,Jambul02,Jambul1}. We note that charge carrier dynamics in
branched polymers have been recently studied in \ci{Chsol} by considering
charged solitons. The scope of our study can be considered in the same context
as those of the
Refs.\ci{Chernyak31,Chernyak34,Chernyak35,Chernyak36,Chernyak37,Chernyak39,Chernyak42,Chernyak43}.
Moreover, Chernyak and co-authors also consider branched polymer as a graph,
although in our model it is considered as quantum graphs, that allows us to use
there well developed quantum graph theory. However, unlike these researches,
where wide aspects of steady state excitons are considered, our study is
focused on energy spectrum and transport of single exciton in branched
conducting polymers.

An advantage of modeling branched structures in terms of metric graphs is the
fact that it allows to describe the structure as one-dimensional, or quasi-one
dimensional system. Here we consider star shaped branched polymers, where the
exciton dynamics is described by the Schr\"odinger equation on metric star
graph. This paper is organized as follows. In the next section we give
formulation of the problem together with the description of the model. In
Section III we apply the model to the exciton dynamics in star-shaped
conducting polymers. Finally, Section IV presents some concluding remarks.

\begin{figure}[t!]
\includegraphics[width=80mm]{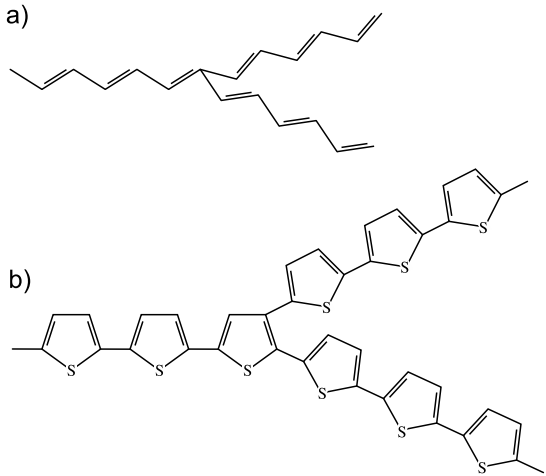}
\caption{Sketch of  branched conducting polymers.} \label{fig1}
\end{figure}

\section{Steady state excitons in a star-branched polymer}

Excitons in conducting polymers are the main charge carriers in photophysical
processes and organic optoelectronic devices. Developing realistic models
exciton migration is of crucial importance for engineering novel functional
materials and optimization of existing ones. Apart from the exciton transport,
description of steady state excitons allows to understand basic factors playing
important role in exciton-lattice, exciton-phonon and other interactions.
Different models for exciton dynamics in conjugated polymers have been proposed
so far in the literature (see, e.g., Refs.\ci{Exc1,Exc2,Exc3,Exc4,Exc5}). For
conducting polymers, due to their quasi-one dimensional and periodic structure,
a polymer chain can be considered as a one-dimensional lattice of monomers.
Therefore, most of the models describing excitons in conducting polymers are
the 1D models. This allows, e.g., to use 1D tight-binding approach, to
calculate band structure and charge carrier migration. Such approach was
proposed first in \ci{Rice}. Here we consider dynamics of excitons in a
branched conducting polymers that consist of three polymer chains, which are
connected to each other at single monomer (see, Fig.1). The model we will use
in this paper is close to that proposed by Abe in the Ref.\ci{Exc2} and assumes
neglecting of electrons-phonon interactions, considers lattice as rigid and
Coulomb interaction between the electron is assumed to be weak compared to that
of between the electron and the hole. Also, to avoid appearing of the infinite
binding energy, we include a cut-off length into the Coulomb potential
\ci{Exc2}. Then the attractive Coulomb potential between the electron and hole
can be written as $$V(x) =-\frac{1}{|x +\xi|}.$$

\begin{figure}[t!]
\includegraphics[width=65mm]{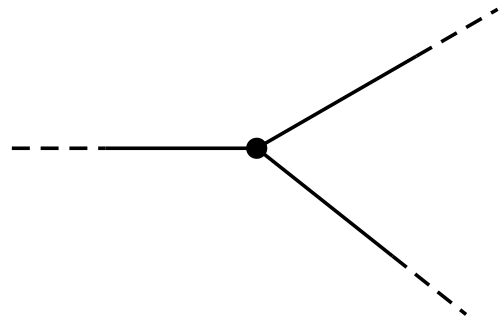}
\caption{Basic star graph.} \label{fig2}
\end{figure}

Here we consider branched polymer in the form of Y-junction (see, Fig.1).
However, our approach can be applied for arbitrary branching topology. Such
system can be mapped onto the basic star graph presented in Fig.2. We assume
that the size of the exciton is much smaller than that of polymer branch and
both electron and hole belong to the same branch. Then the dynamics of
electron-hole pair can be effectively reduced to one-particle description in
terms of center-of-mass coordinates (see Appendix for details of separation of
variables in the two-particle Schr\"odinger equation). The stationary
Schr\"odinger equation on metric star graph describing exciton in branched
polymer (on each branch of the graph presented in Fig.2 and in the units
$\hbar=e=1$) can be written as \be
\left(\frac{p^2}{2\mu}-\frac{1}{|x+\xi_j|}\right)\psi_j(x)=E_b\psi_j(x),
\lab{lse1}\ee where $p=-i\frac{d}{dx},$ $E_b$ is the binding energy for
electron-hole pair, $j$ is the number of branch. Coulomb potential describes
electron-hole interaction and $\mu$ is the reduced mass given by
$$
\frac{1}{\mu}=\frac{1}{m_e}+\frac{1}{m_h},
$$

To solve Eq.\re{lse1}, one needs to impose the boundary conditions at the
branching points (vertices) of the graph. Here we impose them as the continuity
of the wave function at the branching point: \be \psi_1(0)=\psi_2(0)=\psi_3(0)
\lab{vbc0101}\ee and the Kirchhoff rule, which is given by \be
\sum\limits_{j=1}^3{\frac{d}{dx}\psi_j(x=0)}=0. \lab{vbc0202}\ee

As the analytical solution of Eq.\re{lse1} is not possible, one needs to solve
it numerically for the boundary conditions \re{vbc0101} and \re{vbc0202}. To do
this, we expand the wave function $\psi_j$ in terms of the complete set of
eigenvalues of star graph: \be \psi_j(x) =\sum c_n\phi_{j,n}(x), \lab{exp1} \ee
where $\phi_{j,n}(x)$ are the eigenvalues of the unperturbed quantum star graph
given by the following Schr\"odinger equation \ci{Keating}: \be
-\frac12\frac{d^2}{dx^2}\phi_j =\varepsilon\phi_j \ee with the boundary
conditions
$$
\phi_1(0)=\phi_2(0)=\phi_3(0)
$$
and current conservation
$$
\sum\limits_{j=1}^3{\frac{d}{dx}\phi_j(x=0)}=0,
$$
and the Dirichlet boundary conditions at the branch ends given by
$$
\phi_1(L_1)=\phi_2(L_2)=\phi_3(L_3) =0.
$$

Explicitly, the eigenfunctions $\phi_{j,n}$ can be written as
\ci{Keating}

\be \phi_{j,n}(x)=\frac{B_n}{\sin{(k_n L_j)}} \sin{(k_n (L_j-x))},
 \ee
where
$$
B_n=\left[\sum\limits_j{(L_j+\sin{(2 k_n L_j)})\sin^{-2}{(k_n
L_j)}}/2\right]^{-1/2}.
$$
Eigenvalues, $\varepsilon$ can be found from the following secular
equation:
$$
\sum_{j=1}^{3}\tan^{-1}\sqrt{2\varepsilon}L_j =0.
$$

Inserting Eq.\re{exp1} into \re{lse1} and multiplying both sides
to $\phi_{j,m}^*$ and integrating by taking into account the
normalization conditions given by
$$
\int \phi_{j,m}^*\phi_{j,n} dx =\delta_{mn},
$$
the eigenvalues, $E_n$ can be found by diagonalizing the matrix $W_{mn}$, which
is given as \be W_{mn} =(\varepsilon_n\delta_{mn} + V_{mn}), \ee with
$\delta_{mn}$ being the Kronecker symbol and
$$
V_{mn} =-\sum_{j=1}^3\int\limits_0^{L_j}{\frac{\phi_{j,m}^*\phi_{j,n}}{|x+\xi_j|}dx}.
$$

In the following numerical calculations we choose $\xi_1=\xi_2=\xi_3 =1$. Fig.3
presents plot of the first 20 energy levels of a steady state exciton on a
branched conducting polymer as a function of the parameter, $\alpha$ given by
the relation $L_j=\alpha l_j, \ j=1,2,3$. The plot shows that as longer the
length of branch, as smaller the binding energy electron and hole in polymer.

\begin{figure}[t!]
\includegraphics[width=90mm]{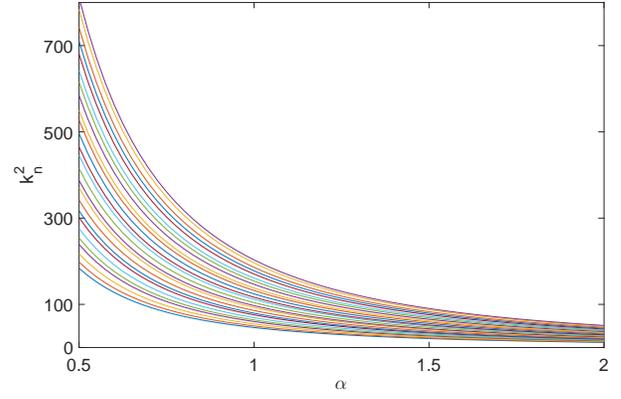}
\caption{The first 20 energy levels of a steady state exciton on a
branched conducting polymer vs the parameter $\alpha$, which is
defined from the relation $L_j=\alpha l_j, \ j=1,2,3.$}
\label{fig3}
\end{figure}

Important characteristics of excitons migrating in branched conducting polymers
is the exciton current at each branch. The total current density is given by
$$
J(x,t) =J_1(x,t) +J_2(x,t) +J_3(x,t),
$$
where
$$
J_j(x,t)=\frac{i}{2}\left[\Psi_j(x,t)\frac{\partial\Psi_j^*(x,t)}{\partial
x}-\frac{\partial\Psi_j(x,t)}{\partial x}\Psi_j^*(x,t)\right],
$$
is the current density on each branch and
$$
\Psi_j(x,t)=\underset{n}{\sum}G_ne^{-ik_n^2t}\psi_j(x,k_n),
$$
where coefficients $G_n$ are found from the initial condition for Eq.\re{lse1}.
The current can be found by integrating current density over $x$.

In Fig.4 the evolution of the current density in time and space is plotted for
the initial condition given in the form of the Gaussian wave packet
\begin{equation*}
\Psi^I(x)=\left(\sqrt{2\pi}
\sigma_0\right)^{-1/2}\exp{\left(-\frac{(x-x_0)^2}{4\sigma_0^2}+ip_0x\right)}.
\end{equation*}
compactly supported in the first branch and centered around $x_0=L_1/2$ with an
average initial momentum $p_0=-5$ and width $\sigma_0=1$. In numerical
calculations we choose lengths of the branches as $L_1=6.62, L_2=7.06$ and
$L_3=6.36$.

\begin{figure}[t!]
\includegraphics[width=95mm]{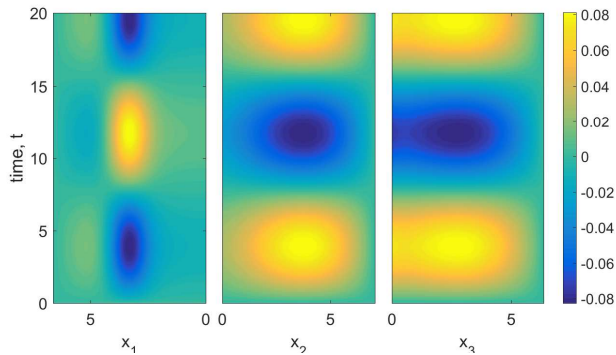}
\caption{Spatio-temporal evolution of the current density for the initial wave
packet given in the form of a Gaussian with parameters $x_0=L_1/2$, $p_0=-5$,
$\sigma_0=1$. Lengths of the branches are $L_1=6.62, L_2=7.06$ and $L_3=6.36$.}
\label{fig4}
\end{figure}

\section{Transport of excitons}

Important issue in modeling of exciton dynamics conducting polymers is
migration of excitons along the polymer chain. In case of branched conducting
polymer, dynamics become richer due to the transmission or reflection of
exciton at the branching point. Here we consider this problem by modeling
exciton dynamics in terms of the time-dependent Schr\"odinger equation on
metric graph, which is given by (in the system of units $\mu=\hbar=1$)
\begin{equation}
i\frac{\partial\psi_j}{\partial t}+\frac{1}{2}\frac{d\psi_j}{dx^2}+
\left(\frac{1}{x+\xi_j}\right)\psi_j =0,\,\, x\in b_j \lab{lse2}
\end{equation}
where $\mu$ is the reduced mass, $\xi_j > 0$ is the cutoff length and $j$ is
the branch number. The boundary conditions are imposed as continuity of the wave function weight
\begin{equation}
\alpha_1\psi_1(0,t)=\alpha_2\psi_2(0,t)=\alpha_3\psi_3(0,t),\label{bc1}
\end{equation}
and current conservation
\begin{equation}
\frac{1}{\alpha_1}\partial_x\psi_1(x=0,t)=\frac{1}{\alpha_2}\partial_x\psi_2(x=0,t)+\frac{1}{\alpha_3}\partial_x\psi_3(x=0,t)\label{bc2},
\end{equation}
where $\alpha_j,\,j=1,2,3$ are real constants. Dirichlet boundary conditions
are imposed at the end of each branch:
\begin{equation}
\psi_j(x=L_j,t)=0,\,j=1,2,3.\label{bc3}
\end{equation}
When these boundary conditions are applied to branched conducting polymers, $\alpha_j$ can be chosen as hopping constants of the polymer ``lattice'', or any physical characteristic of a branch, e.g., its conductance.

\begin{figure}[t!]
\includegraphics[width=85mm]{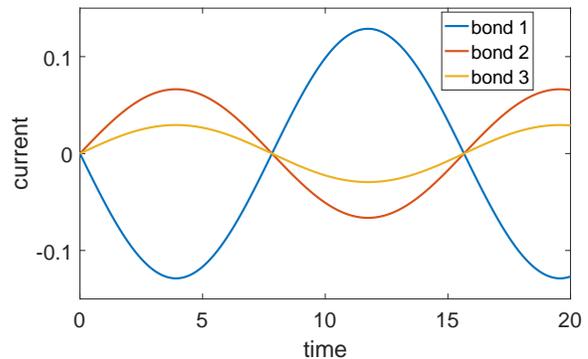}
\caption{Time-dependence of current on each branch for the case shown in
Fig.~4.} \label{fig5}
\end{figure}

Here we use concept of so-called transparent boundary conditions (TBC) for
quantum graphs, which was introduced recently in the Refs.\ci{Jambul,Jambul02}.
Unlike the scattering matrix based approach widely used in quantum mechanics,
this novel approach allows to introduce reflectionless propagation of the waves
in terms of boundary conditions for the wave equation. Although explicit form
of these boundary conditions are very complicated and numerical solution of the
problem requires using highly accurate and stable discretization scheme,
application of the approach to quantum graphs allows considerable
simplification. Namely, as it was shown in \ci{Jambul,Jambul02}, under certain
constraints, transparent boundary conditions become equivalent to usual
continuity and Kirchhoff rules given by Eqs. \re{bc1} and \re{bc2}.  Therefore,
the method proposed in \ci{Jambul,Jambul02} may become powerful approach for
solving the problem of reflectionless wave propagation in branched structures
modeled in terms of quantum graphs.

It was shown in \ci{Jambul,Jambul02} that the vertex boundary conditions given
by Eqs.\re{bc1} -\re{bc3} become equivalent to the transparent boundary
conditions, i.e., provide reflectionless transmission of the wave through the
branching point, when parameters $\alpha_j$ fulfill the following sum rule: \be
\frac{1}{\alpha_1^2}=\frac{1}{\alpha_2^2}+\frac{1}{\alpha_3^2} \lab{sumrule}\ee

To check this, time evolution of the profile of the exciton's wave function, initially chosen as a right traveling Gaussian wave packet
$$
\psi^I(x)=(2\pi)^{-1/4}\exp(30ix-(x+5)^2/4),
$$
has been simulated. The Crank-Nicolson finite difference scheme
with the space discretization $\Delta x=0.0067$ and the time step
$\Delta t=1\cdot 10^{-5}$ has been used in this calculations. The
result of this numerical experiment is plotted in Fig.~6,
for the values of BC parameters
$\alpha_1=\sqrt{2/3}$, $\alpha_2=1$, $\alpha_3=\sqrt{2}$.
Reflectionless transmission of the exciton through the polymer
branching point can be seen from this plot.

\begin{figure}[t!]
\includegraphics[width=87mm]{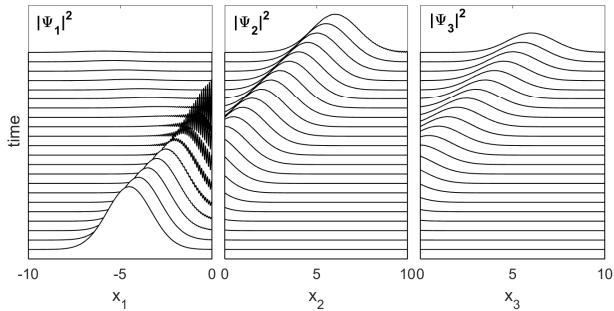}
\caption{The profile of the exciton's wave function on a branched conducting
polymer for the values of $\alpha_j$ fulfilling the sum rule in
Eq.~\eqref{sumrule}: $\alpha_1=\sqrt{2/3}$, $\alpha_2=1$,
$\alpha_3=\sqrt{2}$.} \label{fig6}
\end{figure}

\section{Conclusions}
We studied steady states and migration of exciton in branched conducting
polymers. A model based on the use of quantum graph concept, described by the
linear Schr\"odinger equation on metric graphs is applied. The main focus is
given to calculation of the energy spectra of steady state exciton and to the
transport along the polymer chains accompanied the transmission (reflection) of
exciton at the branching points. Combining the quantum graphs and transparent
boundary conditions concepts, constraints providing the reflectionless
transmission of exciton through the branching point is derived in the form of a
simple sum rule. Numerical computations showing such transmission are provided
in addition to the analytical results. The model proposed in this work can be
extended to the case of more complicated branching topologies. Also, it can be
applied for modeling of exciton migration in polymer based thin film organic
solar cells, where polymer chains packed on the cell create complicated
branched structures.

\section*{Appendix: Two particle system on a quantum graph}

Remarkable feature of quantum-mechanical two particle motion is the fact that
variables in the Schr\"odinger equation can be separated in center of mass and
inter-distance variables. This allows one to reduce two-particle motion
effectively to one-particle description. Here we will show that similar
reducing is possible for two-particle motion on quantum graphs. Consider the
star graph with three branches $b_j$, for which coordinates $x_{1j},\,x_{2j}$ are
assigned. Choosing the origin of coordinates at the vertex, 0, for each branch
$b_j$ we put $x_{1j},\,x_{2j}\in[0,L_j]$. In what follows, we use the shorthand
notation $\psi_j(x_1,x_2)$ for $\psi_j(x_{1j},x_{2j})$, where $x_1,\,x_2$ are
the coordinates on the branch $j$ to which the component $\psi_j$ refers.

Then assuming that both particles are located in the same branch and the distance
between the particles are much shorter than the length of the branch, one can
write the Schr\"odinger equation describing such two particle system on each
branch of graph, $b_j\sim[0;L_j]$  as
\begin{equation}
i\hbar\frac{\partial \psi_j}{\partial
t}=-\frac{\hbar^2}{2m_1}\frac{\partial^2\psi_j}{\partial
x_{1}^2}-\frac{\hbar^2}{2m_2}\frac{\partial^2\psi_j}{\partial
x_{2}^2}+U_j(x_{1}-x_{2})\psi_j,\label{eq1}
\end{equation}
where $U_j(x_{1}-x_{2})$ is the potential of interaction between
the particles. Furthermore, we introduce the relative coordinate
$x=x_{1}-x_{2}$ and the center-of-mass coordinate
$X=\frac{m_1x_{1}+m_2x_{2}}{m_1+m_2}$. In terms of these new
coordinates,  $x$ and $X$  in Eq. (\ref{eq1}) can be rewritten as
\begin{equation}
i\hbar\frac{\partial \psi_j}{\partial
t}=-\frac{\hbar^2}{2\mu}\frac{\partial^2\psi_j}{\partial
x^2}-\frac{\hbar^2}{2(m_1+m_2)}\frac{\partial^2\psi_j}{\partial
X^2}+U_j(x)\psi_j,\label{eq2}
\end{equation}
where $\mu=\frac{m_1m_2}{m_1+m_2}$. Separating variables in Eq.(\ref{eq2}) and
using the substitution
\begin{equation}
\psi_j(x,X,t)=\psi_{jM}(X)\psi_{j\mu}(x)e^{-i\hbar E
t},\label{eq3}
\end{equation}
we get
\begin{eqnarray}
-\frac{\hbar^2}{2\mu}\frac{\partial^2\psi_{j\mu}}{\partial
x^2}+U_j(x)\psi_{j\mu}=E_b\psi_{j\mu},\nonumber\\
\frac{\hbar^2}{2(m_1+m_2)}\frac{\partial^2\psi_{jM}}{\partial
X^2}+\hbar^2E\psi_{jM}=E_b\psi_{jM}.\label{eq4}
\end{eqnarray}

Thus center of mass and relative (inter-distance) motions can be separated  when electron and hole belong in the same
branch.  However, this does not automatically apply to the case of transmission of exciton through the branching area. In that case, depending on the distance between electron and hole, they may appear in different branches. Therefore, to avoid such a situation and provide separation of variables in such case, one needs to make additional assumptions. Here we assume that the size of the branching area is comparable or larger that that of exciton.

\acknowledgements{This work is supported by joint grant of the Ministry for Innovation Development of Uzbekistan and the Federal Ministry of Education and Research (BMBF) of Germany (Ref. No. M/UZ-GER-06/2016 (UZB-007)).}


\begin{thebibliography}{99}

\bibitem{MGH} R.J. Kline, M.D. McGehee, J. Macromolecular Sci. C, {\bf 46} 27 (2006).
\bibitem{Wallace} D.S. Wallacet A.M. Stonehams, W. Hayest, A.J. Fishertt and A. Testas, J. Phys.: Condens. Matter {\bf 3} 3905 (1991).
\bibitem{Kumar} D. Kumar, R. C. Sharma, Eur. Polym. J., {\bf 34} 1053 (1998).
\bibitem{Salaneck} W.R. Salaneck, R.H. Friend, J.L. BreHdas, Phys. Rep.,  {\bf 319} 231 (1999).
\bibitem{Heeger1} A.J. Heeger and R. Pethig, Phil. Trans. R. Soc. Lond. A {\bf 314} 17 (1985).
\bibitem{Heeger2} A.J. Heeger, Rev. Mod.Phys. {\bf 73} 681 (2001).
\bibitem{Heeger3} A.J. Heeger, S. Kivelson, J.R. Schrieffer, W.-P. Su, Rev. Mod.Phys. {\bf 60} 781 (1988).

\bibitem{Chsol1} K. Maki, Synth. Met., {\bf 9} 185 (1984).
\bibitem{Chsol2} L. Rothberg, T. M. Jedju, S. Etemad, G. L. Baker, Phys. Rev. Lett. {\bf 57} 3229 (1982).
\bibitem{Chsol3} M. Kuwabara, S. Abe, and Y. Ono, Synth. Met., {\bf 85} 1109 (1997).
\bibitem{Chsol4} P.B. Miranda, D. Moses, A.J. Heeger, Y.W. Park,  Phys. Rev. B,   {\bf 66} 125202 (2002).
\bibitem{Chsol5} A.J. Heeger, S. Kivelson, J.R. Schrieffer, W.-P. Su, Rev. Mod.Phys. {\bf 60} 781 (1988).
\bibitem{Chsol6} S. Brazovskii, Solid State Sci. {\bf 10} 1786 (2008).

\bibitem{SSH} W.P. Su, Schrieffer, A.J. Heeger,  Phys. Rev. Lett., {\bf 42} 1698 (1979).
\bibitem{TDFT} L. Bernasconi,  J.Phys. Chem. Lett., {\bf 6} 908 (2015).
\bibitem{PPP} M. Sasai, H. Fukutome, Prog. Theor. Phys., {\bf 79}  61 (1988).
\bibitem{TDHF} S. Suhai, J. Chem. Phys., {\bf 73} 3843 (1980).

\bibitem{Rice} M.J. Rice, Yu.N. Gartstein, Phys. Rev. Lett. {\bf 73} 2504 (1994).
\bibitem{Braz1} S. Brazovskii, N. Kirova, A.R. Bishop,  Opt. Mat., {\bf 9} 465 (1998).
\bibitem{Braz02} D. Moses, J. Wang, A.J. Heeger, N. Kirova, S. Brazovski,  Synth. Met., {\bf 119} 503 (2001).
\bibitem{Braz2} N. Kirova, S. Brazovskii, Thin Solid Films, {\bf 403} 419 (2002).
\bibitem{Braz3} N. Kirova, S. Brazovskii, Synth. Met., {\bf 141} 139 (2004).
\bibitem{Braz4} N. Kirova, S. Brazovskii, Current Appl. Phys., {\bf 4} 473 (2004).
\bibitem{Kobrak} M. N. Kobrak,  E. R. Bittner, J. Chem. Phys., {\bf 112} 5399 (2000).
\bibitem{Bittner1} H. Tamura,  E.R. Bittner, I. Burghardt, J. Chem. Phys. {\bf  126},021103 (2007).
\bibitem{Bittner2} H. Tamura, J.G.S. Ramon, E.R. Bittner, I. Burghardt, Phys. Rev. Lett. B {\bf  100}, 107402 (2008).
\bibitem{Bittner3} W. Barford, E.R. Bittner, A. Ward, J. Phys. Chem. {\bf  116}, 10319 (2012).

\bibitem{BCP1} M. Fujii, K. Ari and K. Yoshino,  J. Electrochem. Soc., {\bf 140} 7, (1993).
\bibitem{BCP2} J. Jurkiewicz and A. Krzywicki  Phys. Lett. B {\bf 392}, 29 (1997).
\bibitem{BCP3} K. Inoue, Prog. Polym. Sci. {\bf 25}, 453 (2000).
\bibitem{BCP5} L. Dai, B. Winkler, L. Dong, L. Tong and A. W. H. Mau, Adv. Mater., {\bf 13},  12-13 (2001).
\bibitem{BCP6} D. T. Wu, Synth. Met., {\bf 126}, 289 (2002).
\bibitem{BCP7} C. Gao and D. Yan Prog, Polym. Sci. {\bf 29}, 183 (2004).
\bibitem{BCP8} F. Hua and E. Ruckenstein, Macromolecules {\bf 38}, 888 (2005).
\bibitem{BCP9} R.-H. Lee, W.-Sh. Chen, and Y.-Y. Wang, Thin Solid Films, {\bf 517}, 5747 (2009)
\bibitem{BCP10} A. Hirao and H.-S. Yoo, Polymer Journal, {\bf 43}, 2 (2011).
\bibitem{BCP11} L. R. Hutchings Macromolecules,  {\bf 45}, 5621 (2012).
\bibitem{BCP12} J. Ohshita, Y. Tominaga, D. Tanaka, T. Mizumo, Y. Fujita, Y. Kunugi, J. Org. Chem., {\bf 50}, 736 (2013).
\bibitem{BCP13}G. V. Otrokhov, O. V. Morozova, I. S. Vasileva, G. P. Shumakovich, E. A. Zaitseva, M. E. Khlupova, and A. I. Yaropolov,
 Biochemistry, {\bf 78}, 1539 (2013).
\bibitem{BCP14} M. Goll, A. Ruff, E. Muks, F. Goerigk, B. Omiecienski, I. Ruff, R. C. Gonzalez-Cano, J. T. L. Navarrete, M. C. R. Delgado and S. Ludwigs,
 Beilstein J. Org. Chem., {\bf 11}, 335 (2015).
\bibitem{BCP15} H Higginbotham, K. Karon, P. Ledwon and Display and Przemyslaw Data, Imaging, {\bf 2}, 207 (2017).
\bibitem{BCP16} T. Soganci, O. Gumusay, H. C. Soyleyici , M. Ak, Polymer 134, 187 (2018).

\bibitem{Chernyak31} Ch. Wu, S.V. Malinin, S. Tretiak, V.Y. Chernyak, Nature Physics, \textbf{2}, 631635 (2006).
\bibitem{Chernyak34} Ch. Wu, S.V. Malinin, S. Tretiak, and V.Y. Chernyak, J. Chem. Phys., \textbf{129}, 174111 (2008).
\bibitem{Chernyak35} Ch. Wu, S.V. Malinin, S. Tretiak, and V.Y. Chernyak, J. Chem. Phys., \textbf{129}, 174112 (2008).
\bibitem{Chernyak36} Ch. Wu, S.V. Malinin, S. Tretiak, and V.Y. Chernyak, J. Chem. Phys., \textbf{129}, 174113 (2008).
\bibitem{Chernyak37} H. Li, S.V. Malinin, S. Tretiak, and V.Y. Chernyak, J. Chem. Phys., \textbf{132}, 124103 (2010).
\bibitem{Chernyak39} H. Li, Ch. Wu, S.V. Malinin, S. Tretiak, V.Y. Chernyak, J. Phys. Chem. B, \textbf{115}(18), pp.5465-5475 (2011).
\bibitem{Chernyak42} H. Li, S.V. Malinin, S. Tretiak, and V.Y. Chernyak, J. Chem. Phys., \textbf{139}, 064109 (2013).
\bibitem{Chernyak43} H. Li, M.J. Catanzaro, S. Tretiak, V.Y. Chernyak, J. Phys. Chem. Lett., \textbf{5}(4), pp.641-647 (2014).

\bibitem{Kost} V.Kostrykin and R.Schrader  J. Phys. A: Math. Gen. {\bf 32} 595 (1999)
\bibitem{Uzy1} T.Kottos and U.Smilansky, Ann.Phys., {\bf 76} 274 (1999).
\bibitem{Kuchment04} P.Kuchment, Waves in Random Media, {\bf 14} S107 (2004).
\bibitem{Uzy2} S.Gnutzmann and U.Smilansky, Adv.Phys. {\bf 55} 527 (2006).
\bibitem{Exner15} P.Exner and H.Kovarik, {\it Quantum waveguides.} (Springer, 2015).
\bibitem{Keating} J.P.Keating, Contemp. Math.,   {\bf 415}, 191 (2006).
\bibitem{Uzy3} S.Gnutzmann, J.P.Keating, F. Piotet, Ann.Phys., {\bf 325} 2595 (2010).
\bibitem{KarimBdG} K.K.Sabirov, J.Yusupov, D. Jumanazarov, D. Matrasulov, Phys.Lett. A, {\bf 382}, 2856 (2018).
\bibitem{PTSQGR}  D.U. Matrasulov, J.R. Yusupov and K.K. Sabirov, J. Phys. A, {\bf 52}, 155302 (2019).
\bibitem{Jambul} J.R. Yusupov, K.K. Sabirov, M. Ehrhardt and D.U. Matrasulov, Phys. Lett. A, {\bf 383}, 2382 (2019).
\bibitem{Jambul02} Aripov M.M., Sabirov K.K., Yusupov J.R., {\it Nanosystems: physics, chemistry, mathematics}, {\bf 10}(5), P.501-602
(2019).
\bibitem{Jambul1} J.R. Yusupov, K.K. Sabirov, M. Ehrhardt and D.U.
Matrasulov, Phys. Rev. E, {\bf 100}, 032204 (2019).
\bibitem{Chsol} D. Babajanov, H. Matyoqubov and D. Matrasulov, J. Chem. Phys.,  \textbf{149}, 164908 (2018).

\bibitem{Exc1} Th G. Pedersen, Phys. Rev. B {\bf 69} 075207 (2004).
\bibitem{Exc2} Sh. Abe, J. Phys. Soc. Jpn. {\bf 58} 62 (1989).
\bibitem{Exc3} Sh. Abe, et.el.  Phys. Rev. B {\bf 45} 9432 (1992).
\bibitem{Exc4} Sh. Abe, J. Yu,  W. P. Su,   Phys. Rev. B {\bf 45} 8264 (1992).
\bibitem{Exc5} A. V. Nenashev,   Phys. Rev. B {\bf 84} 035210 (2011).



\end{thebibliography}
\end{document}